\documentclass[journal]{IEEEtran}
\usepackage{amsmath}
\usepackage{amsthm}
\usepackage{amsfonts}
\usepackage{amssymb}
\usepackage[pdftex]{graphicx}
\renewcommand{\vec}[1]{\mathbf{#1}}

\newcommand{\Roff}{R_{\mathrm{off}}}
\newcommand{\Ron}{R_{\mathrm{on}}}
\newcommand{\Loff}{L_{\mathrm{off}}}
\newcommand{\Lon}{L_{\mathrm{on}}}
\newcommand{\Voff}{V_{\mathrm{off}}}
\newcommand{\Von}{V_{\mathrm{on}}}

\usepackage{hyperref}
\usepackage{amsmath}
\usepackage{amssymb}
\usepackage{amsfonts}
\usepackage{graphicx}
\renewcommand{\vec}[1]{\mathbf{#1}}

\newtheorem{theorem}{Theorem}[section]

\usepackage{cite}

\ifCLASSINFOpdf
\else
\fi
\begin{document}
%
\title{Uniform and Piece-wise Uniform Fields in Memristor Models}
%
%
%

\author{Ella~Gale \thanks{E. Gale is with the International Center for Unconventional Computing and 
Bristol Robotics Laboratory, Frenchay Campus, Coldharbour Lane, Bristol,
BS16 1QY UK e-mail: ella.gale@brl.ac.uk }}%
\IEEEpeerreviewmaketitle

\maketitle


\begin{abstract}
The Strukov model was the phenomenological model that accompanied the announcement of the first recognised physical instantiation of the memristor and, as such, it has been widely used. This model described the motion of a boundary, $w$, between two types of inter-converting material, $\Roff$ and $\Ron$, seemingly under a uniform field across the entire device. In fact, what was intended was a field with a discontinuity at $w$, that was uniform between $0<x<w$. In this paper we show that the discontinuity is required for the Strukov model derivation to be completed, and thus the derivation as given does not describe a situation with a uniform field across the entire device. The discontinuity can be described as a Heaviside function, $H$, located on $w$, for which there are three common single-valued approximations for $H(w)$. The Strukov model as intended includes an approximation for the Heaviside function (the field is taken to be the same as that across the $\Ron$ part of the device). We compare 
approximations and give solutions. We then extend the description of the field to a more-realistic continuously varying sigmoidal transition between two uniform fields and demonstrate that the centro-symmetric approximation model (taking the field as being the average of the fields across $\Ron$ and $\Roff$) is a better single-point model of that situation: the other two approximations over or underestimate the field.
\end{abstract}

\section{Introduction}







\IEEEPARstart{M}{emristors} are electronic components, often nanoscale in size, that act as resistors with a memory. This combination of functionality and small size has led to the suggestion that memristors are a potential route to increasing computational complexity in terms of Moore's Law~\cite{278}. Memristor theory has been successfully applied to describe the operation of synapses~\cite{84} and other components of neurons~\cite{247,248}, as well as the processes of learning in snails~\cite{51} and amoeba~\cite{3}. Therefore, memristors may become vital components in attempts to create brain-like (neuromorphic) computers that are capable of learning (see for example~\cite{239} and \cite{DavidJ1}) and possibly higher-level functions, such as intelligence.

There are four fundamental circuit properties which describe a circuit's operation: the electrical potential difference, $V$, the electronic current, $I$, the magnetic flux, $\varphi$ and the charge, $q$. Three pairs of relationships define the operation of the first three fundamental circuit elements: the resistor ($R=V/I$); the capacitor ($C=q/V$); the inductor ($I=\varphi/V$). A fourth was added in 1971, when Chua predicted the existence of a device that would relate $q$ to $\varphi$, the memristor\cite{14}. Because $I$ and $V$ are time differentials of $q$ and $\varphi$, memristors produce distinctive non-linear $I$-$V$ curves that have three important features~\cite{279}: (1) hysteresis (memory); (2) zero current at zero voltage; and (3) A.C. frequency dependence (the size of the hysteresis is related to frequency and it shrinks to nothing above critical frequency). The memristor concept has been  generalised to memristive systems~\cite{84}. The memristor has only one state variable whereas a memristive 
system can be a function of more, and memristive systems have been used to model systems from across the sciences from an alternative circuit model of the neuron~\cite{84,247,249} to thermistors~\cite{93}.

Between 1971 and 2008, no one who had read Chua's work was able to make a memristor and the idea languished in the drawer of theoretical curiosities. However, over this time period, many experimentalists had reported `anomalous' I-V curves with variations on a pinched hysteresis loop, such as the first report of a memristor I-V curve in TiO$_2$ in 1968~\cite{183}, the creation of the PEO-PANI organic based memristive system~\cite{12,5} and many others (see ~\cite{313,Review1,338} for recent reviews), mostly Resistive Random Access Memory (ReRAM) devices~\cite{119,236,280}. Strukov et al were the first to finally unite the idea of the memristor with a physical example, by describing a working memristor~\cite{15} complete with a phenomenological model for its operation and references to Chua's theoretical work.  

In terms of understanding the Strukov memristor, much work~\cite{348,320,2,46,306,328,327} has been done on modelling memristor properties with SPICE and similar simulation tools (see~\cite{Review1,346,345} for overviews), but there have been few theoretical models. The first with material and device properties was Strukov et al's phenomenological model which describes the operation of their device~\cite{15} and this model has been shown to be a good approximation for the real-world device behaviour, despite not addressing the non-linearity of dopant drift and not implicitly containing the effect of the boundaries. These known issues have been well-addressed through the usage of windowing functions~\cite{2,46,306}. Two theories have extended it, ~\cite{83} demonstrated a SPICE model with the initial state as a variable and ~\cite{224} extended the model using Bernoulli formulation, introducing extra parameters to get a measure of the device hysteresis. Other models have been applied to this device, such as 
an analytical one based on electromagnetic modelling of vacancy current~\cite{F0c} and one based on fitting the experimental data with 8 parameters~\cite{220} with simulational improvements~\cite{307,305,308,310}, however a large proportion of current memristor simulation work is based on the original Strukov model or one of the extensions to it. 

The validity of the Strukov model has been questioned in many papers. The lack of a demonstrable magnetic flux term, which is expected from Chua's constitutive relation has led to questions as to whether the Strukov memristor is a Chua memristor\cite{142}. In response, it has been suggested that the magnetic flux may be a theoretical construct and not related to a material property~\cite{119} and/or that it is the non-linear $I$-$V$ relationship which defines the memristor~\cite{15}, a definition which widens the field of memristors to include, for example, all ReRAM devices~\cite{119}, or that the relevant magnetic flux is that related to the vacancy current~\cite{F0c}. Further questions have been asked by the electrochemical community~\cite{94}. Even the corresponding author of the original paper~\cite{15} has asserted that the model is `not for science'~\cite{289}. Regardless of this, the Strukov model has been used as the basis of a great deal of work, due to its simplicity and the fact that it is based 
on a real world device that few scientists have access to. Many use the model as it is to test out their design ideas and many have worked to improve the known shortcomings (linear ionic motion and unphysical edge effects at the boundary) of the model and some have based their more in-depth analyses on it. For this reason, we believe that it is crucial that the memristor community be made aware of exactly what the derivation is and the issues caused by the wide-spread misunderstanding that has arisen from a lack of grammatical clarity in the original published model. In this paper we undertake a critical examination the model and the assumptions made during its creation to formulate a better version for wide-spread use.

This paper is structured in the following manner. First we will describe Strukov's derivation as it appears to be from reading the paper as published and present solutions for this model. Then we will describe the derivation as it actually was done, discuss the approximations made and their meaning, and offer alternatives. 

\section{The Strukov Model of the Strukov Memristor}

\subsection{Chua's Definition of the Memristor}

There are four circuit measurables: $V$, $I$, $q$ and $\varphi$. The definitions

\begin{equation}
q(t) \equiv \int_{- \infty}^{t} I(\tau) d\tau
\label{eq:QuIT}
\end{equation}
and
\begin{equation}
\varphi(t) \equiv \int_{- \infty}^{t} V(\tau) d\tau
\end{equation} 
relate charge to current and voltage to flux. They also widen the description of charge from a quantity stored by a capacitor to total amount of charge that has passed through the circuit. Similarly the flux is time integral of the voltage applied over time rather than the quantity stored by an inductor. Thus these quantities are relevant to circuits without such devices in them. 

Between four measurables, there are six pairs of interactions. Two are given by the definitions above, another three are the constitutive relations of the resistor, inductor and capacitor described in the introduction. Chua's ground-breaking contribution was to realize that there was a constitutive relationship missing, one which would define a device that relates charge to flux. He realised that this device would act in a passive, (i.e. non-storing) manner (like a resistor) and would give rise to a pinched hysteresis loop, and this hysteresis suggested the device would have memory, hence the name memristor, a contraction of memory-resistor. This constitutive relationship~\cite{14} he gave as:
\begin{equation}
M(q) \equiv \frac{d \varphi(q)}{d q}
\label{eqn:Chua}
\end{equation}
where $M$ is the memristance, a time-varying instantaneous resistance, which relates voltage, $V$, to current, $I$, in the following manner
\begin{equation}
V(t) = M(q(t)) I(t)\: .
\label{eq:V=MI}
\end{equation}
The time dependence of the memristance demonstrates that there will be a time- or frequency-based effect on the response of the device. If the voltage is varied too quickly for the device to respond the memristive behaviour collapses to ohmic conduction. Note that the time-variance is entirely due to the fact that $q$ is a function of $t$, the memristance will not change with time if $q$ does not change. These equations are for a charge-controlled memristor, it is trivial to work out those for a flux-controlled device. The equations reproduced in these sections are the definitions given in Chua's original theoretical work~\cite{14} (the description is expanded to memristive systems in later works~\cite{84}), to our knowledge Chua has never given a description of what the function $M$ is or how it arises from material properties (this is because $M$ is based on circuit theory and not materials science).

\subsection{Strukov's Derivation of the Strukov Memristor Model. \label{sec:Stru}}

At the time of publication, Strukov's ground-breaking paper was the first example of a working memristor which excited much interest and founded a novel field and increased interest in existing fields, such as ReRAM. Since then, as it was discovered that other memristor systems had been fabricated, the impact of that paper on memristor theory has been the usefulness of the phenomenological model described within it. By providing a description for $M$, theorists were able to model memristor devices and make progress in memristor science, even whilst many of them did not have real devices to test their ideas on. The model was found to fit other memristor systems~\cite{71} and the modelling of the boundary has been improved (i.e. by using windowing functions and more realistic models for oxygen ion transport). However, there were two errors in the derivation, which has led to confusion over the magnetic flux and the claim (now experimentally proved false) that memristors had to be nanoscale~\cite{15}. In this 
section we shall go through the derivation from paper~\cite{15} in detail to highlight these issues and demonstrate why the derivation is incorrect.

\subsubsection{The Details of Strukov's Derivation}

\begin{figure}[!t]
\includegraphics[width=3.5in]{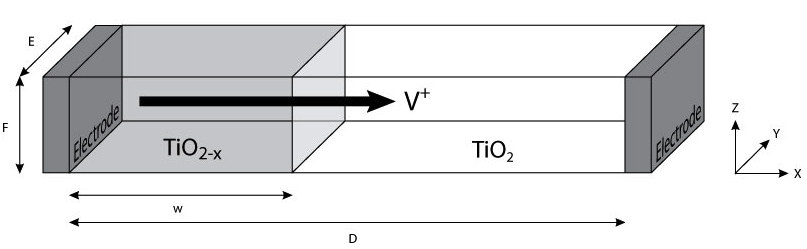}
\caption{\label{fig1} The Strukov memristor. The shaded area is doped low-resistance titanium dioxide, the unshaded area is stoichiometric high-resistance titanium dioxide. Vacancy, V$^{+}$, movement through the material is shown by the arrow. $w$ marks the boundary between the two forms of titanium dioxide. The limits of the titanium dioxide layer in the $y$ and $z$ directions are $E$ and $F$. As the vacancies move to the right along the $x$ axis, the magnetic $\vec{B}$ field associated with them curls around in an anti-clockwise direction (not shown) and thus the surfaces that cross magnetic field lines are those in the $x$-$y$ and $x$-$z$ planes.}
\end{figure}

The Strukov memristor~\cite{15} (also often called the H.P. memristor after the funders of the work) consists of a layer of titanium dioxide of thickness $D$ sandwiched between two platinum electrodes of width $E$ and $F$ as shown in figure~\ref{fig1}. The titanium dioxide layer contains lattice defects caused by missing oxygen atoms; these oxygen vacancies act as $n$-type dopants and the distance they have drifted through the memristor is given by $w$, where $0<w<D$. The resistivity of the stoichiometric, un-doped TiO$_2$ is higher than that of doped, non-stoichiometric TiO$_{(2-x)}$. Interconversion between the doped and un-doped forms, caused by the drift of oxygen vacancies, changes the total resistance, $R$, of the memristor and produces the pinched hysteresis loop in the $I$-$V$ curve.


To deal with the `simplest possible case'~\cite{15}, the authors make two assumptions:
\begin{enumerate}
 \item `ohmic electronic conductance', i.e. $V(t) = R(t) I(t)$, $R$ may vary with time but at each instant in time, the current is wholly determined by the voltage and resistance; 
\item `linear ionic drift in a uniform field with average ion mobility $\mu_v$'.
\end{enumerate}

Because the TiO$_2$ can inter-convert to TiO$_{(2-x)}$, with no volume change, the device can be modelled as two space-conserving resistors, one with resistance $\Roff$ and one with resistance $\Ron$. The resistors are space-conserving because TiO$_{(2-x)}$ is made from the TiO$_2$ and vice-versa, thus they have made a zeroth assumption that matter is not created or destroyed in this system (not stated in the paper, although perhaps too trivial to require stating). Strukov et al started with the space-conserving variable resistors and gave the following formula for how the resistance changes as equation 5 in~\cite{15}
\begin{equation}
V(t) = \left( \Ron \frac{w(t)}{D} + \Roff \left( 1 - \frac{w(t)}{D}\right)\right) I(t) ,
\label{eq:SWR}
\end{equation}
where $\frac{w(t)}{D}$ serves to give the fraction of the material which is the doped resistor and $\left( 1 - \frac{w(t)}{D}\right)$ is the undoped resistor. This is a simple model of two variable resistor with a runner between (see figure 2A in~\cite{15}), mathematically it is a proportional mixing of two resistors, which gives rise to a continuum change of resistance. 

From equation~\ref{eq:SWR} and assumption 1 (ohmic conduction) we can write an expression for the total resistance in the system, $R(t)$, as:
\begin{equation}
 R(t) = \Ron \frac{w(t)}{D} + \Roff \left( 1 - \frac{w(t)}{D}\right) \; .
\label{eq:Rt}
\end{equation}

Strukov et al concern themselves with the speed of the moving boundary, $\frac{d w(t)}{dt}$. This is modelled in relation to the average drift velocity of the oxygen dopants, $\vec{s}_{v}$. This arises from assumption 2: if we have linear ionic drift, then the ions move (on average) at the same speed across the entire device. Therefore because the boundary is the measure of the furthest reach of those ions, the boundary must move at the drift speed (because if the travelling front of ions moved (on average) faster or slower than the bulk average it would contradict the assertion that we have linear ionic drift). As the average drift speed is the scalar part of the average ionic drift velocity, we have 
\begin{equation}
 \frac{d w(t)}{dt} = |\vec{s}_{v}| \; .
\label{eq:svwt}
\end{equation}

Assumption 2 also states that the ions are drifting in a uniform electric field $\vec{L}$ with an average ionic mobility $\mu_v$, which gives us (from writing out assumption 2, which is a definition of drift velocity):
\begin{equation}
 \vec{s}_{v} = \mu_v \vec{L} \;.
\label{eq:v_d}
\end{equation}
A uniform electric field is given by the voltage across the titanium dioxide divided by the thickness of that material:
\begin{equation}
 L = \frac{V}{D} \; .
\label{eq:L}
\end{equation}

So, they set $\frac{d w(t)}{d t} = | \vec{s}_{v}|$ and substitute equation~\ref{eq:L} into~\ref{eq:v_d} to get
\begin{equation}
\frac{d w(t)}{d t} = \mu_v \frac{V(t)}{D} \: ,
\label{eq:orig}
\end{equation}
which is not explicitly given in the paper~\cite{15} but is described in words in the derivation and arises naturally from the assumptions. 

The authors actually report (as equation 6 in~\cite{15}) 
\begin{equation}
\frac{d w(t)}{d t} = \mu_v \frac{I R_{\mathrm{on}} }{D} \: ,
\label{eq:SW6}
\end{equation}
where assumption 1 (ohmic electronic conductance) has been used to substitute for $V$.

They then integrate both sides 
\begin{equation}
\int \frac{d w(t)}{d t} d t = \int \mu_v \frac{I(t) R_{\mathrm{on}} }{D} d t \: ,
\end{equation}
and make use of the definition of charge, given in equation~\ref{eq:QuIT} to get 
\begin{equation}
w(t) = \mu_v \frac{\Ron q(t)}{D} \; ,
\label{eq:Sw7}
\end{equation}
which is equation 7 in~\cite{15}. 

Equation~\ref{eq:Sw7} is then substituted into equation~\ref{eq:SWR} to give
\begin{equation}
V(t) = \left( \frac{\Ron^{2} \mu_v q(t)}{D^{2}} + \Roff - \frac{\Roff \Ron \mu_v q(t)}{D^{2}} \right) I(t) \; .
\label{eq:Swq}
\end{equation}

They compare equation~\ref{eq:Swq} to Chua's constitutive relation given in equation~\ref{eq:V=MI} and conclude that $R(t)$ (the terms in the brackets) is the memristance, $M(q)$, where the time dependence of $M$ arises entirely from $q(t)$. 

They state that:
\begin{equation}
 \Ron \ll \Roff \; ,
\label{eq:A3}
\end{equation}
which as $\Ron \sim 1$ and $\Roff \sim 160$ in their system, is not unreasonable. We shall thus take equation~\ref{eq:A3} as assumption 3. Therefore, as $\Ron$ is over 100 times smaller than $\Roff$ they drop the $\Ron^{2}$ term (to make the equation simpler). Thus they report
\begin{equation}
M(q) = \Roff \left( 1 - \frac{\mu_v \Ron q(t)}{D^2} \right) \: .
\label{eq:SWM}
\end{equation}

\subsubsection{Consequences of the Strukov derivation}

As a result of equation~\ref{eq:SWM} several conclusions were made that have led to misunderstandings about memristors. 

Equation~\ref{eq:Rt} is theoretically one-dimensional as it is dependent only on $w$. This arises from treating the memristor as two-space conserving resistors with a slider $w$ to mix the relative amount of $\Roff$ and $\Ron$. The model is also spatially one-dimensional because the amount of 3-D memristor material in the doped and un-doped state is approximated by only the proportion of the thickness $D$ in that state. This is not a bad assumption given that (because we assume that the volume occupied by the memristive material does not change as a result of doping) the length and width of memristive material is the same for the doped and undoped memristive material. An alternative formulation would be to replace $\Ron$ and $\Roff$ terms with the resistivity of the two types of material (if we know it), and the volume that material occupies using the definition of resistivity.

This unrecognised assumption leads to an error in the conclusion -- from examination of Strukov's expression for $M$, the conclusion was made that memristance depends only on $D$~\cite{15}, meaning that memristance is spatially one-dimensional. However, $M$ depends only on $D$ and not any other dimensions of the device because no other spatial dimensions were included in the model at the start (for simplicity). Thus to conclude that $M$ depends only on $D$ is to make the error of drawing a conclusion that was an un-expressed assumption. 

This error is relatively minor, indeed to my knowledge the only published work to remark on it is~\cite{255} (despite the existence of devices which cannot be described as 1D memristors, namely the `plastic memristor' which has 2 perpendicular currents~\cite{12,45}), but the compound error associated with $\frac{1}{D^{2}}$ has had a demonstrable effect on the field of memristors. Strukov et al suggested that for appreciable memristance to be measured, $D$ must be small, ideally nanoscale, to make $\frac{1}{D^{2}}$ (and thus the difference between the fully switched on and fully switched off resistances) large. Thus it was suggested that memristors could not be fabricated until nanoscale ($D \sim 10^{-9}$m) film technology existed, because memristance was a nanoscale phenomenon~\cite{15}, `essentially unobservable at the millimetre scale'~\cite{330}. There have since been several experiments that contradict this viewpoint~\cite{M0,29,236} and old memristor papers have come to light~\cite{119,236,316,313}. 
However, the only reason that the memristance didn't depend on $E$ or $F$ was because they were purposely not included at the start. (This realisation does suggest the intriguing possibility of designing memristors with different properties based on their shape).

Another historical problem with equation~\ref{eq:SWM} is the question of where the magnetic flux is. Because Chua's definition (see equation~\ref{eqn:Chua}) included magnetic flux~\cite{14} it was expected that there should be an equation that related Strukov's memristance to a magnetic flux. However, the authors state that the ``magnetic field does not play an explicit role in the mechanism of memristance''~\cite{15}. They concluded that memristance was just a theoretical concept and it was only the non-linear relation between the integrals of voltage and current that defined a memristor. 

We will now go on to deconstruct the Strukov derivation as published. We shall approach the problem in a mathematical way as: `words can be ambiguous and interpreted (or twisted) in different ways', so `when a question arises about the specific meaning of a concept, we must go back to the defining equations'~\cite{334}.

\subsection{Mathematical Critique of the Strukov Derivation as  published in~\cite{15})}

The problem with the derivation arises from the substitution between equations~\ref{eq:orig} and~\ref{eq:SW6} where $I R$ is substituted for $V$. 

\subsubsection{Assumptions}

For simplicity, we list the assumptions used by Strukov et al below:

\begin{itemize}
 \item 0. Matter is conserved in this system
 \item 1. The system is instantaneously ohmic: $V(t) = R(t) I(t)$
 \item 2. The ions undergo linear ionic transport in a uniform field with an average ion mobility of $\mu_v$
 \item 3. $\Ron \ll \Roff$
\end{itemize}

We also have other physical facts about the system, such as the value of the electron mobility, $\mu_{e}$, that $w$ varies between 0 and $D$, that all the resistances are more than zero, that $D$ is non-zero and so on, which gives us a fourth assumption:

\begin{itemize}
 \item 4. `$w$ specifies the distribution of dopants in the device' and `is bounded between 0 and D' and is `proportional to the charge $q$ that passes through the device'~\cite{15}. And from the values given: $\Ron > 0$, $\Roff > 0$, $D > 0$, $0 < w < D$, $\mu_e > 0$, $\mu_v > 0$.
 \item Note that in this section, we are assuming that there is a uniform field, $\vec{L}$, between 0 and $D$ associated with the total voltage drop $V$ (this is based on assumption 2).
\end{itemize}

\subsubsection{The published derivation does not describe a uniform field across the entire device}

\begin{theorem}
 The substitution of $\Ron$ into equation~\ref{eq:orig} that leads to equation~\ref{eq:SW6} is incorrect
\label{th:R}
\end{theorem}

\begin{IEEEproof}
  Assumption 1 is $V(t) = R(t) I(t)$ where $R(t)$ is the total resistance and $I(t)$ is the total current. Therefore, when substituting for $V$ in equation~\ref{eq:SW6}, the resulting equation should be
  \begin{equation}
  \frac{d w(t)}{d t} = \mu_v \frac{I R }{D} \: .
  \label{eq:SW6Correct}
  \end{equation} 
\end{IEEEproof}

If equation~\ref{eq:SW6} were correct, it leads to either a description of a non-memristor or a contradiction (see theorem~\ref{th:Ron}), if we consider the resistance, and is an incorrect description of $\frac{d w}{d t}$, if we consider the current, see theorem~\ref{th:I} in the appendix. Taking equation~\ref{eq:SW6Correct} as a starting point and following the derivation through (as given in theorem~\ref{th:Rt} in the appendix) we get:

\begin{equation}
 w(t) = \frac{D \Roff \left(\mathrm{e}^{\frac{\mu_v \left( \Ron - \Roff \right) q(t)}{D^2}} \right)}{\left( \Ron - \Roff \right)} \: ,
\label{eq:WBetter}
\end{equation}

for $w$ and
	
	\begin{equation}
	R(t) = M(q) = \Roff \mathrm{e}^{-\frac{\mu_v \left( \Roff - \Ron \right) q(t)}{D^2}} \; ,
	\label{eq:SwSoln}
	\end{equation}
	
	for $M(q)$. As this is not the equation presented in~\cite{15} it demonstrates that the authors were not using a uniform field across the entire device.

\section{The Actual Strukov Derivation}

The single uniform field interpretation of the paper~\cite{15} is incorrect. This fact was made known to the author in a communication~\cite{289} with the corresponding author and Nature about paper~\cite{15}, so the relevant points will be quoted here:

\begin{quote}
 ``The field is not constant across the device as you assume [V(t)/D].  In our very simple model, there is a discontinuity in the field at the point w. The voltage drop inside the doped region, which will limit the rate of motion, is RON * w(t)/D * i(t) [i.e. the resistance of the doped region times the current].  Therefor, the field is this voltage drop divided by the length, or RON * w(t)/D * i(t) / w(t) = RON*w(t)/D * i(t), as in the paper.''
\end{quote}

Thus, a clearer writing of the authors' intention for paper~\cite{15} is: 
\begin{quote}
`The application of an external bias $v(t)$ across the device will move the boundary between the two regions by causing the charged dopants to drift. For the [simplest] case of ohmic electronic conduction and linear ionic drift in a uniform field \textit{across each region} with average ion mobility
$\mu_v$ ~\textit{within region $R_{on}$, i.e. between 0 and $w$}...'
\end{quote}
where the suggested additions are in italics and deletions in brackets.

Thus, assumptions 2 and 5 should be: 



\begin{itemize}
 \item 2. The ions undergo linear ionic transport an average ion mobility of $\mu_v$
\item 5. The field across the device is uniform between 0 and $w$, there is a second field between $w$ and $D$ and $w$ is located on the discontinuity~\footnote{The form of the second field was not discussed and we take the position of assuming that it is also uniform, although it can have any form and the following discussion still stands}. 
\end{itemize}

With these  changed assumptions, the derivation in the paper~\cite{15} can be completed as written, as long as the same approximation is chosen for the field at the discontinuity, $L(w)$.

\section{The Hidden Approximation  Within the Actual Model}

\begin{figure}[t!]

 \centering
 \includegraphics[width=3.5in]{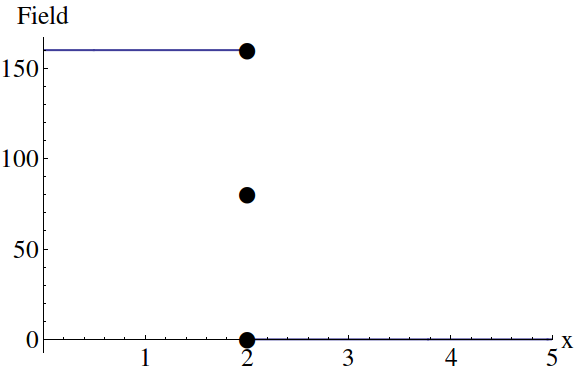}
 \caption{The Heaviside step function (with values for the Strukov memristor) in blue, with possible approximations for $H(w)$ as black dots, note $w=2$ in this example.}
 \label{fig:StepFunction}
\end{figure}

Assumption 5 causes some issues. By placing a discontinuity at $w$, the field at that point is undefined, which prevents the derivation of an analytical answer. The authors get around this by making an implicit approximation of the field strength at this point, $L(w)$. The total field across the device is equivalent to the Heaviside function of the form:

\begin{equation}
L(x,t) = (\Loff-\Lon) H[x-w(t)] + \Lon
\end{equation}

which describes how the field changes across the device (along the $x$ axis) through time, see figure~\ref{fig:StepFunction}. Note that $L$ is the total field across the device, $L(w)$ is its value at $w$, $\Lon$ is the field across the doped part of the device (that part described by a resistor of $\Ron$ and $\Loff$ is the field across the undoped part of the device, namely that part described by a resistor $\Roff$. The Heaviside function is defined as

\begin{eqnarray}
 H[y] & = & 0, y < 0  \; ;\\
 H[y] & = & 1, y > 0 \; ;\\
\end{eqnarray}
and so acts as a switch to turn the feild from $\Lon$ to $\Loff$. Although the Heaviside function is not defined at zero, three approximations are commonly made:
\begin{enumerate}
\item $H[0] = 0;$
\item $H[0] = 1;$
\item $H[0] = \frac{1}{2}$.
\end{enumerate}
Approximation 1 is used when the `0' state is most important~\footnote{This is equivalent to placing $w$ at the edge of the $\Ron$ volume, adjacent to the discontinuity}, approximation 2 is used when the `1' state is most important, and approximation 3 is usually thought to be the best because is it rotationally symmetric and takes into account the states either side of $H[0]$. 
For the field described above, we get the three approximation options:
\begin{enumerate}
 \item $H[x-w(t)] = \Lon;$
  \item $H[x-w(t)] = \Loff;$
  \item $H[x-w(t)] = \frac{(\alpha \Lon + \beta \Loff)}{(\alpha + \beta)} = L_w$,
\end{enumerate}
where we have used a more complicated form to account for non-linearities were a skewed `discontinuity' considered. In the simplest case, which we shall consider from here on in, $\alpha = \beta = 1.$ 

The values for the fields can be easily calculated, using the equations above, these are:

\begin{eqnarray}
 \Lon & = & \frac{\Von}{w} = \frac{1}{w}\frac{w}{D} \Ron I(t)\\ 
      & = & \frac{\Ron I(t)}{D} ~\label{eq:Lon}\\
 \Loff & = & \frac{\Voff}{(D-w)} = \frac{1}{(D-w)}\left(1-\frac{w}{D} \right) \Roff I(t)\\
      & = & \frac{\Roff I(t)}{D} ~\label{eq:Loff}\\
 L_w & = & \frac{\left( \Roff + \Ron \right) I(t)}{2 D} ~\label{eq:Lw}\\
\end{eqnarray}

This means the field is independent of $w$, which makes sense we are calculating the field at $w$ and what we have is a moving discontinuity, the local field is always the same. In reality, we might expect the field at $w$ to alter as $w$ moves across the device, for this, we can make $\alpha$ and $\beta$ $x$-dependent. i.e. $\alpha(x)$ and $\beta(x)$, and apply window functions, but here we will continue with the simplest cases where the material is assumed to inter-convert with the same ease all along the $x$ axis of the device. 

We can now follow through the derivation and get solutions for each approximation.

\subsection{Solutions for the Three Approximations}

\subsubsection{Approximation 1: $L=L_{on}$}

For completeness, the equations for $L(w) \approx \Lon$ are (as in~\cite{15}):

\begin{equation}
 w(t) = \frac{\mu_v}{D} \Ron q(t) \;,
\label{eq:L_-w}
\end{equation}

and 

\begin{eqnarray}
R(t) & = & \frac{\mu_v}{D^2}  \Ron^2  q(t) + \Roff \\
     &   & - \frac{\mu_v}{2 D^2}  \Ron \Roff q(t) \; .\\
 \label{eq:L_on-R}
\end{eqnarray}

\subsubsection{Approximation 2: $L=\Loff$}

If we were to approximate $L(w)$ as $\Loff$ the solutions are:

\begin{equation}
 w(t) = \frac{\mu_v}{D} \Roff q(t) \;,
\label{eq:L_off-w}
\end{equation}

and 

\begin{eqnarray}
R(t) & = & \frac{\mu_v}{D^2} \Ron \Roff q(t) + \Roff \\
     &   & - \frac{\mu_v}{D^2} \Roff^2 \Ron q(t) \; .\\
 \label{eq:L_off-R}
\end{eqnarray}

\subsubsection{Approximation 3: $L = L_{w}$}

If we set $\mu_v L_w$ equal to $\frac{dw}{dt}$ and follow the equations above we get:

\begin{equation}
 w(t) = \frac{\mu_v}{2 D} \left( \Roff+\Ron \right) q(t) \;,
\label{eq:L_w-w}
\end{equation}

and 

\begin{eqnarray}
R(t) & = & \frac{\mu_v}{2 D^2} \left( \Ron^2 + \Roff \right) q(t) + \Roff \\
     &   & - \frac{\mu_v}{2 D^2} \left( \Roff^2 + \Ron \Roff \right) q(t) \; .\\
 \label{eq:L_w-R}
\end{eqnarray}

\begin{figure}[t!]
 \centering
 \includegraphics[width=3.5in]{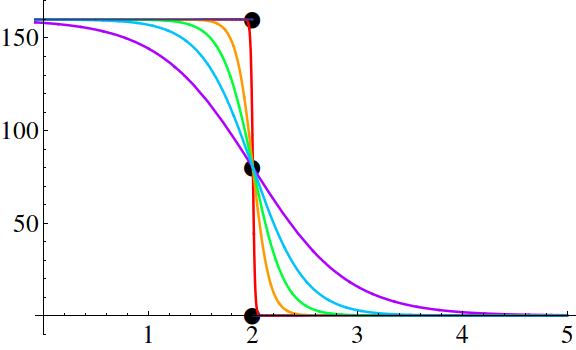}
 \caption{Approximations for the Heaviside Step function as given in equation~\ref{eq:SmoothedOut}. Key: blue: Heaviside function; red: t=0.01; orange: t=0.08; green: t=0.15; cyan: t=0.25 purple: t=0.5}
 \label{fig:SmoothedOut}
\end{figure}

\subsection{Comparison of Approximations}

As paper~\cite{15} is a short letter, the choice of approximation for the Heaviside function, or even that an approximation is made at this point, is not discussed. From the solution presented in~\cite{15} and the correspondence, we know that the authors chose to approximate the system using approximation 1 ($L=\Lon$), which can be argued as being appropriate as the doped part of the device is the most important if we are modelling the movement of ions, in fact in the correspondence (quoted above) the authors stated that they chose this approximation because they assumed the boundary is limited by the movement of ions behind it. In strict numerical terms, however this choice of approximation contradicts assumption 3: since, equations~\ref{eq:Lon} and ~\ref{eq:Loff} implies that $\Loff >> \Lon$ (from assumption 3). To approximate the field at $w$, $L(w)$, by $\Lon$ implies that $\Lon$ is the more important (i.e. larger feild): $\Lon > \Loff$ which contradicts assumption 3: $\Loff < \Lon \bot \Loff >> \Lon$.

This leaves two approximations, $L(w) = \Loff$ and $L(w) = L_w$. Using $\Loff$ as the approximation does not introduce any mathematical contradictions. Nonetheless, we posit that $L_w$ is the best approximation for three reasons: 
\begin{enumerate}
 \item $L_w$ is centro-symmetric around $w$ and thus won't introduce any asymmetry in modelling the charging and discharging the device;
 \item If we consider making the model more physical by smoothing the Heaviside function out to a sigmoidal curve, the field at point $w$ will equal $L_w$ (assuming no skew), see figure~\ref{fig:SmoothedOut};
 \item Physically, we would expect a vacancy located at $w$ to be affected by both fields: the one behind it ($\Lon$) and the one ahead of it ($\Loff$). 
\end{enumerate}
Although any of the three approximations may be used, having the field at $w$ being between the $\Lon$ and $\Loff$ seems the most sensible. Using $L_w$ implies that the boundary moves at a different speed to ions in the doped part of the device (because if the ions are assumed to be undergoing linear ionic transport, their velocity is only due to the field, and thus the drift velocity ions far away from the boundary is proportional to $\Lon$, those at the boundary have a drift velocity proportional to $\frac{1}{2} \left( \Lon + \Loff \right)$, which is greater than $\Lon$). Specifically, the ions move faster at the boundary due to higher field in the off part of the device having an effect on the ions at the boundary.  

Another reason to prefer $L_w$ over $\Loff$ comes from analysing the system using the numbers in paper ($V=\pm 1V$) for a Strukov memristor~\cite{15}. Approximation 2, $\Loff$, requires 25,157.2C of charge to fully switch the memristor compared to 0.322011C for approximation 3, (using $L_w$), and 0.161006C for approximation 1 (using $\Lon$). The large charge requirement is equivalent to that required by a 25,157.2F super-capacitor and suggests that this would  be an unphysical model. All solutions give us a resistance that decreases linearly with charge, i.e. $M(q) \propto q$. $L_w$ decreases and half the rate of $\Lon$, $\Loff$ decreases over 10,000 times slower with charge. Using the values of the resistances in~\cite{15}, approximation 3 gives a speed of $w$ is 80.5 times faster, than approximation 1, and approximation 2 gives a speed 160 times faster than approximation 1.


The argument here has been entirely mathematical in nature. Obviously, a real discontinuity of this form would have a huge effect on the local vacancy movement, introducing significant non-linearities in the ionic transport at $w$ (contradicting the assumption of `linear ionic transport'). For a discussion of the electrochemical issues with a field of this form, see~\cite{294}. The assumption has also been made that the concentration of vacancies is constant across the volume $w \Ron$. Chemically, as the number of missing oxygen ions is only 3\% and as all materials have defects, we cannot expect that the flow of oxygen vacancies through the material is limited by the supply in $w \Ron$. There is no chemical reason why the local concentration of vacancies might not increase (or decrease) as the boundary travels slower (or faster) than the bulk ions behind it, and these ions at the boundary would only be limited by those behind if there were some sort of `communication' between them, i.e. if the 
vacancies were monomer residues in a polymer or hopping dopants which could not get too close due to crystal structure considerations. The model calculates the average drift velocity, and there is no reason why the boundary could not be moving faster or slower than bulk average, and, mathematically we have shown it should move at a different speed, and the physics of the device suggest that the ions at the boundary will feel an effect of $\Loff$ and behave differently.

\subsection{A Smoother Model}

Obviously, the model presented here is a simplification of a real system. In nature a system of this type would not exist, instead the discontinuity would be smoothed out as a sigmoidal curve connecting $\Lon$ and $\Loff$. We define $S$ as the width of this sigmoidal curve (i.e. the distance over which the field changes from approximatable as uniformly $\Lon$ and approximatable as uniformly $\Loff$), see figure~\ref{fig:SmoothedOut}.

\subsection{Advantages of the Models}

There are problems with modelling the device with a discontinuity in the middle. It has been argued that because the Strukov model is based on bulk properties (i.e. $\mu_v$, $\vec{s}$) it is inherently a macroscopic model unsuited to modelling a nanoscale device~\cite{294}, although this does not preclude its use in macroscopic memristors (such as~\cite{M0}). We suggest, however, that a simple model can work well enough to prototype device behaviours for use in circuit simulations.

Molecular dynamics simulations of the ionic vacancies would not work with the discontinuity because as the ions moved across the discontinuous field they would get an unphysical impulse pushing them forward. A smooth sigmoidal transition would be easier to model and give more physical results. 

By putting a discontinuity in the middle of the field, we have introduced an edge. There has been much work with window functions on improving the action of the Strukov model at the edges at the ends of the device (i.e. the limits of $w$: 0 and $D$), so it is suggested that these could be applied to the edge in the middle of the device.

Even acknowledging these issues, is it possible to use a simple model with a discontinuous field as a useful approximation to the real physics? The Heaviside part of the field can be smeared out to cover a transition region, where the field would have a sigmoidal curve (which could be skewed) between $\Loff$ and $\Lon$. The centre point of this field would be (assuming no skewness) $L_w$ and by setting the simpler Strukov model's $w$ at that point, it can be a simple approximation to the real situation. Thus, the model could be made more realistic by replacing the piece-wise uniform field with a more complex form. 

There are a few different approximations to the Heaviside step function which can be used to model the real situation, that of a continuously-varying field with two bulk values, a good choice~\footnote{Analytically, this function is a good choice because it is relatively simple, centro-symmetric and easy to understand (in that $t$ is a measure of boundary region). For our argument, we only need an example analytical model, and suggest that simulationists may be able to pick a more computationally efficient and easier-to-numerically calculate function} is:
\begin{equation}
 \bar{H}(0) = \frac{1}{1+ e^{-\frac{(w-x)}{t}}} \; ,
\label{eq:SmoothedOut}
\end{equation}
where $t$ is a measure of the extent of the boundary region (defined by that part of the device where the field is not $\Loff$ or $\Lon$) and $t = \frac{1}{2}S$, $\bar{H}$ is an approximation for the Heaviside function.  Examples of possible solutions for this field is shown in figure~\ref{fig:SmoothedOut}, the functions all have the same value as the Heaviside function for assumption 3 (the centre point), for assumption 1 and 2 the approximations introduce an error on calculated field. From equation~\ref{eq:SmoothedOut}, we can see that the smoothed out function has the same value as the Heaviside function when the exponential term is zero, this happens when $w = x$ which corresponds to the middle dot in figure~\ref{fig:SmoothedOut} (assumption 3) and when $x=w\pm t$, which corresponds to points in figure~\ref{fig:SmoothedOut} where the curves approach $\Loff$ and $\Lon$, and $\pm t$ is the error in $x$ from using assumption 1 or 2 to approximate the field at $w$. The error in the field $L(w)$ for the 
situation described in fig.~\ref{fig:SmoothedOut} is $\pm \frac{1}{2}\Loff-\Lon$, 
which is a factor of 79.5 for the Strukov memristor described in~\cite{15}. For a situation like that presented in~\ref{fig:SmoothedOut}, assumption 3 is the best single point assumption for the field at $w$.



\section{Conclusions}

In this paper we have highlighted a subtle, but key point, in understanding the derivation of the first phenomenological model of memristance, specifically that the field is not uniform and assuming a uniform field leads to a different model to that published. 
We have then analysed the model using a discontinuous field with $w$ located on this discontinuity and considered the three possible approximations for the field at $w$. Of these three solutions, the approximation that takes the field at $w$ to be half way between the field in the doped and undoped parts of the device is the best and approximates the actual system (as represented by a smooth sigmoidal transistor between $\Lon$ and $\Loff$) well. Choosing one of the other approximations leads to significant differences in the simulations. 

Whilst the use of a discontinuity at this point is not suitable for molecular dynamics simulations, this simplification is usable for modelling the device at a bulk behaviour level. We hope that simulationists and engineers will find this model a useful abstraction. We suggest that there is further work to be done in modelling the device and taking account of the field around $w$.   

\section{Appendix}

In this section, we outline the possibilities for equation~\ref{eq:SW6} to be correct for the situation of a uniform field across the device and demonstrate that would lead to contradictions. This shows that the derivation as published cannot be applied to a uniform field across the device and that we must consider a discontinuous field with the discontinuity located on $w$.

\begin{theorem}
If equation~\ref{eq:SW6} were correct, it leads to either a description of a non-memristor or a contradiction
\label{th:Ron}
 \end{theorem}

\begin{IEEEproof}
  For equation~\ref{eq:SW6} to be correct,
  \begin{equation}
  R = \Ron \\
  \Rightarrow \Ron \frac{w(t)}{D} + \Roff \left( 1 - \frac{w(t)}{D}\right)  = \Ron \; ,
  \end{equation}
  from substituting for $R$.

  There are two ways this can be possible. 

  The first way is if $w(t) = D \; \forall \; t$, this implies that $w$ can not be dependent on $t$ i.e. it does not change,
  \begin{equation}
  \therefore \frac{d w(t)}{d t} = 0
  \label{eq:StuckOn}
  \end{equation}
  which, because 
  \begin{equation}
  \mu_v, R, D \neq 0 \Rightarrow I(t) = 0 \; \forall \; t. \:  \IEEEQED
  \label{eq:Unpowered}
  \end{equation} 

  Equation~\ref{eq:StuckOn} describes the system when it is stuck at the minimum resistance, which is equivalent to a resistor of resistance $\Ron$ and no longer fits the definition of the memristor. Equation~\ref{eq:Unpowered} describes an un-powered device (that can never be turned on), thus it is also not a memristor.

  The second way for $R=\Ron$ is if $\Roff = \Ron$, then
  \begin{equation}
  R(t) = \Ron \frac{w(t)}{D} + \Ron \left( 1 - \frac{w(t)}{D}\right) \; ,
  \end{equation}
  which if we rewrite using $x = \frac{w(t)}{D}$ 
  \begin{equation}
  R(t) = x \Ron  +  \left( 1 - x \right) \Ron \; ,
  \end{equation}
  we know that $ 0 \leq w \leq D \Rightarrow 0 \leq x \leq 1$, thus we see that $R(t) = \Ron$. But this involved setting $\Roff$ to $\Ron$ which contradicts assumption 3:
  \begin{equation}
  \Ron = \Roff \perp \Ron \ll \Roff \; .
  \end{equation}
\end{IEEEproof}

\subsubsection{Problems with the current}

\begin{theorem}
 The substitution of $V$ in equation~\ref{eq:orig} by $I R$ is an incorrect expression for $\frac{d w(t)}{d t}$
\label{th:I}
\end{theorem}

\begin{IEEEproof}
 The zeroth assumption says that matter can not be created or destroyed in this system, thus we can derive Kirchhoff's laws. From Kirchhoff's laws, the total measured current ($I$) is a sum of all the currents in the system, specifically the electronic current $i_{e}$ and the ionic vacancy current $i_{v}$:
\begin{equation}
  I(t) = i_{v} + i_{e} \; .
\end{equation}
Thus the right hand side of equation~\ref{eq:SW6} (now we are ignoring the issues with the resistance) should be 
\begin{equation}
 \mu_v \frac{\Ron}{D} \left( i_{v} + i_{e} \right) \; ,
\label{eq:sRHS}
\end{equation}
which is actually a measure of the average drift velocity of all the charge carriers in the system, $\vec{s}$, i.e. $\mu_v \frac{\Ron}{D} \left( i_{v} + i_{e} \right) = s$.

Thus
\begin{equation}
 \mu_v \frac{\Ron}{D} \left( i_{v} + i_{e} \right) = \vec{s} \perp \mu_v \frac{V}{D}  = \vec{s}_{v} ,
\end{equation}
unless $\vec{s}=\vec{s}_{v}$.

From equation~\ref{eq:v_d}, the average drift velocity of all charge carriers can be expressed as
\begin{equation}
\vec{s} = \left[ \frac{n_{v} \mu_{v}}{N} + \frac{n_{e} \mu_{e}}{N} \right] L \; ,
\label{eq:s}
\end{equation}
where $n_{v}$ is the number of oxygen vacancy charge carriers in the system, $n_{e}$ is the number of electron charge carriers in the system and $N$ is the total number of charge carriers given by $N = n_{v} + n_{e}$.

$\vec{s}=\vec{s}_{v}$ iff
\begin{equation}
  \frac{n_{v} \mu_{v}}{N} + \frac{n_{e} \mu_{e}}{N}  = \mu_v \;
\end{equation}
which can happen in two cases:
\begin{enumerate}
 \item if $n_e =0 \Rightarrow N = n_v \Rightarrow \frac{n_{v} \mu_{v}}{n_{v}} = \mu_v$
 \item if $\mu_e = 0 \Rightarrow e$ are not charge carriers, $\Rightarrow N = n_{v}$ .
\end{enumerate}

Case 1 suggests that the memristor device is ionic rather than mixed ionic and electronic. This is not mathematically impossible, but does not fit with the physics of the system we are trying to model. Case 2 contradicts the known values~\footnote{$\mu_e$ in this system is around 1$\times 10^{-6}cm^{-2}V^{-1}s^{-1}$~\cite{124}, $\mu_v$ is around $1\times10^{-10}cm^{-2}V^{-1}s^{-1}$~\cite{15}} of the mobilities: $\mu_e = 0 \perp \mu_e > \mu_v$ and $\mu_e, \mu_v \nless 0$. 
\end{IEEEproof}

This concludes our disproof by exhaustion of the correctness of equation~\ref{eq:SW6}. To demonstrate that the following equations are unsupported, we need to prove that if we replace $\Ron$ with $R$ we cannot derive the expressions for memristance~\ref{eq:SWM}.

\subsubsection{The Corrected Version of Equation~\ref{eq:SW6} cannot be Used to Derive the Expression for Memristance in Equation~\ref{eq:SWM}} 

\begin{theorem}
 Using $V=IR$ to substitute for $V$ in equation~\ref{eq:orig} leads to a different expression for memristance.
\label{th:Rt}
\end{theorem}

\begin{IEEEproof}
  If we substitute $R$ into equation~\ref{eq:SW6} in place of $\Ron$, we would get 
  
  \begin{equation}
  \frac{dw}{dt} = \frac{\mu_v I R}{D} \; .
  \label{eq:SwWithR}
  \end{equation}
  
$w$ is a function of $t$, $I$ is a function of $t$ and $R$, unlike $\Ron$ is a function of $t$ because it is also a function of $w$. To get an expression for $w$ from~\ref{eq:SwWithR} we can write:

  \begin{equation}
  \int \frac{1}{R(w)} d w = \int \frac{\mu_v I(t)}{D} dt
  \label{eq:SwW}
  \end{equation}  
  
which from~\ref{eq:QuIT} and ~\ref{eq:Rt} gives:

	\begin{equation}
	\frac{D \mathrm{ln}[ D \Roff + w \left( \Ron-\Roff \right) ]}{\left( \Ron - \Roff \right)} = \frac{\mu_v q}{D} + C \; .
	\label{eq:wrong}
	\end{equation}
	
	Solving~\ref{eq:wrong} for $w$ gives
	
	\begin{equation}
	w = \frac{ \mathrm{e}^{\mu_v q \left( \Ron - \Roff \right)+C D\left( \Ron - \Roff \right)} - D \Roff}{\left( \Ron - \Roff \right)} \: 
	\label{eq:SwW2}.
	\end{equation}

We know, from the definition of $q$, that $q$ should equal 0 when $w = 0$, solving~\ref{eq:SwW2} for these numbers gives us an expression for the constant of integration:

\begin{equation}
 C = \frac{D \mathrm{ln} \left[ \frac{1}{D \Roff} \right] }{\left(\Ron -\Roff \right)} \; , 
\label{eq:C}
\end{equation}

which when substituted into equation~\ref{eq:SwW2} gives
\begin{equation}
 w(t) = \frac{D \Roff \left(\mathrm{e}^{\frac{\mu_v \left( \Ron - \Roff \right) q(t)}{D^2}} \right)}{\left( \Ron - \Roff \right)} \: 
\label{eq:WBetter}
\end{equation}

	which when substituted into~\ref{eq:Rt} and simplified gives the following expression for $R(t)$ (and hence the memristance as a function of $q$):
	
	\begin{equation}
	R(t) = M(q) = \Roff \mathrm{e}^{-\frac{\mu_v \left( \Roff - \Ron \right) q(t)}{D^2}} \; .
	\label{eq:SwSoln}
	\end{equation}
	
Equation~\ref{eq:SwSoln} is not the same as that presented in the paper. Thus, we have shown that that equation~\ref{eq:SWM} cannot been derived from the steps in~\cite{15}.



\end{IEEEproof}


%
\section*{Acknowledgment}

The author is indebted to Stan Williams for his useful discussions and critiques. E.G. would like to thank Oliver Matthews for useful discussions and editorial advice.



\bibliographystyle{IEEEtran}
\bibliography{UWELit}

\ifCLASSOPTIONcaptionsoff
  \newpage
\fi

\end{document}